\documentclass[reprint,aps,prd,twocolumn,superscriptaddress,showpacs,preprintnumbers,amsmath,amssymb,floatfix,altaffilletter]{revtex4-1}

\usepackage{graphicx}           
\usepackage[mathlines]{lineno}  
\usepackage{xspace}             
\usepackage{array}
\usepackage{multirow}
\usepackage{hyperref}

\newcommand{\etal}{\textit{et al.}\xspace}
\newcommand{\gev}{GeV/$c^2$\xspace}
\newcommand{\co}{$^{60}$Co\xspace}
\newcommand{\cf}{$^{252}$Cf\xspace}
\newcommand{\cs}{$^{137}$Cs\xspace}

\begin{document}


\title{Low-threshold analysis of CDMS shallow-site data}

\affiliation{Department of Physics, Brown University, Providence, RI 02912, USA}
\affiliation{Division of Physics, Mathematics and Astronomy, California Institute of Technology, Pasadena, CA 91125, USA}
\affiliation{Department of Physics, Case Western Reserve University, Cleveland, OH  44106, USA}
\affiliation{Fermi National Accelerator Laboratory, Batavia, IL 60510, USA}
\affiliation{Lawrence Berkeley National Laboratory, Berkeley, CA 94720, USA}
\affiliation{Department of Physics, St.\,Olaf College, Northfield, MN 55057 USA}
\affiliation{Department of Physics, Princeton University, Princeton, NJ 08544, USA}
\affiliation{Department of Physics, Queen's University, Kingston, ON, Canada, K7L 3N6}
\affiliation{Department of Physics, Santa Clara University, Santa Clara, CA 95053, USA}
\affiliation{SLAC National Accelerator Laboratory/KIPAC, Menlo Park, CA 94025, USA}
\affiliation{Department of Physics, Southern Methodist University, Dallas, TX 75275, USA}
\affiliation{Department of Physics, Stanford University, Stanford, CA 94305, USA}
\affiliation{Department of Physics, Syracuse University, Syracuse, NY 13244, USA}
\affiliation{Department of Physics, Texas A \& M University, College Station, TX 77843, USA}
\affiliation{Department of Physics, University of California, Berkeley, CA 94720, USA}
\affiliation{Department of Physics, University of California, Santa Barbara, CA 93106, USA}
\affiliation{Departments of Physics and of Electrical Engineering, University of Colorado Denver, Denver, CO 80217, USA}
\affiliation{Department of Physics, University of Florida, Gainesville, FL 32611, USA}
\affiliation{School of Physics and Astronomy, University of Minnesota, Minneapolis, MN 55455, USA}
\affiliation{Physics Institute, University of Z\"{u}rich, Winterthurerstrasse 190, CH-8057, Switzerland}

\author{D.S.~Akerib} \affiliation{Department of Physics, Case Western Reserve University, Cleveland, OH  44106, USA}
\author{M.J.~Attisha} \affiliation{Department of Physics, Brown University, Providence, RI 02912, USA}
\author{L.~Baudis} \affiliation{Physics Institute, University of Z\"{u}rich, Winterthurerstrasse 190, CH-8057, Switzerland}
\author{D.A.~Bauer} \affiliation{Fermi National Accelerator Laboratory, Batavia, IL 60510, USA}
\author{A.I.~Bolozdynya} \altaffiliation[Now at ]{National Research Nuclear University MEPhI, 115409 Moscow, Russia} \affiliation{Department of Physics, Case Western Reserve University, Cleveland, OH  44106, USA}
\author{P.L.~Brink} \affiliation{SLAC National Accelerator Laboratory/KIPAC, Menlo Park, CA 94025, USA}
\author{R.~Bunker} \thanks{Corresponding author:  bunker@hep.ucsb.edu (R.~Bunker)} \affiliation{Department of Physics, University of California, Santa Barbara, CA 93106, USA}
\author{B.~Cabrera} \affiliation{Department of Physics, Stanford University, Stanford, CA 94305, USA} 
\author{D.O.~Caldwell} \affiliation{Department of Physics, University of California, Santa Barbara, CA 93106, USA} 
\author{C.L.~Chang} \altaffiliation[Now at ]{Enrico Fermi Institute, University of Chicago, Chicago, IL 60637, USA} \affiliation{Department of Physics, Stanford University, Stanford, CA 94305, USA}
\author{R.M.~Clarke} \affiliation{Department of Physics, Stanford University, Stanford, CA 94305, USA}
\author{J.~Cooley} \affiliation{Department of Physics, Southern Methodist University, Dallas, TX 75275, USA} 
\author{M.B.~Crisler} \affiliation{Fermi National Accelerator Laboratory, Batavia, IL 60510, USA}
\author{P.~Cushman} \affiliation{School of Physics and Astronomy, University of Minnesota, Minneapolis, MN 55455, USA} 
\author{F.~DeJongh} \affiliation{Fermi National Accelerator Laboratory, Batavia, IL 60510, USA} 
\author{R.~Dixon} \affiliation{Fermi National Accelerator Laboratory, Batavia, IL 60510, USA}
\author{D.D.~Driscoll} \altaffiliation[Now at ]{Kent State University, Ashtabula Campus, Ashtabula, OH 44004, USA} \affiliation{Department of Physics, Case Western Reserve University, Cleveland, OH  44106, USA}
\author{J.~Filippini} \affiliation{Division of Physics, Mathematics and Astronomy, California Institute of Technology, Pasadena, CA 91125, USA} 
\author{S.~Funkhouser} \affiliation{Department of Physics, University of California, Berkeley, CA 94720, USA}
\author{R.J.~Gaitskell} \affiliation{Department of Physics, Brown University, Providence, RI 02912, USA}
\author{S.R.~Golwala} \affiliation{Division of Physics, Mathematics and Astronomy, California Institute of Technology, Pasadena, CA 91125, USA}
\author{D.~Holmgren} \affiliation{Fermi National Accelerator Laboratory, Batavia, IL 60510, USA} 
\author{L.~Hsu} \affiliation{Fermi National Accelerator Laboratory, Batavia, IL 60510, USA} 
\author{M.E.~Huber} \affiliation{Departments of Physics and of Electrical Engineering, University of Colorado Denver, Denver, CO 80217, USA}
\author{S.~Kamat} \affiliation{Department of Physics, Case Western Reserve University, Cleveland, OH  44106, USA}
\author{R.~Mahapatra} \affiliation{Department of Physics, Texas A \& M University, College Station, TX 77843, USA}
\author{V.~Mandic} \affiliation{School of Physics and Astronomy, University of Minnesota, Minneapolis, MN 55455, USA}
\author{P.~Meunier} \affiliation{Department of Physics, University of California, Berkeley, CA 94720, USA}
\author{N.~Mirabolfathi} \affiliation{Department of Physics, University of California, Berkeley, CA 94720, USA} 
\author{D.~Moore} \affiliation{Division of Physics, Mathematics and Astronomy, California Institute of Technology, Pasadena, CA 91125, USA} 
\author{S.W.~Nam} \altaffiliation[Now at ]{National Institute of Standards and Technology, Boulder, CO 80305, USA} \affiliation{Department of Physics, Stanford University, Stanford, CA 94305, USA}
\author{H.~Nelson} \affiliation{Department of Physics, University of California, Santa Barbara, CA 93106, USA} 
\author{R.W.~Ogburn} \affiliation{Department of Physics, Stanford University, Stanford, CA 94305, USA} 
\author{X.~Qiu} \altaffiliation[Now at ]{Department of Physics, Stanford University, Stanford, CA 94305, USA} \affiliation{School of Physics and Astronomy, University of Minnesota, Minneapolis, MN 55455, USA} 
\author{W.~Rau} \affiliation{Department of Physics, Queen's University, Kingston, ON, Canada, K7L 3N6}
\author{A.~Reisetter} \affiliation{School of Physics and Astronomy, University of Minnesota, Minneapolis, MN 55455, USA} \affiliation{Department of Physics, St.\,Olaf College, Northfield, MN 55057 USA} 
\author{T.~Saab} \affiliation{Department of Physics, University of Florida, Gainesville, FL 32611, USA}
\author{B.~Sadoulet} \affiliation{Lawrence Berkeley National Laboratory, Berkeley, CA 94720, USA} \affiliation{Department of Physics, University of California, Berkeley, CA 94720, USA}
\author{J.~Sander} \affiliation{Department of Physics, Texas A \& M University, College Station, TX 77843, USA}
\author{C.~Savage} \altaffiliation[Now at ]{Oskar Klein Centre for Cosmoparticle Physics, Department of Physics, Stockholm University, SE-106 91 Stockholm, Sweden} \affiliation{Department of Physics, University of California, Santa Barbara, CA 93106, USA}
\author{R.W.~Schnee} \affiliation{Department of Physics, Syracuse University, Syracuse, NY 13244, USA}
\author{D.N.~Seitz} \affiliation{Department of Physics, University of California, Berkeley, CA 94720, USA} 
\author{T.A.~Shutt} \altaffiliation[Now at ]{Department of Physics, Case Western Reserve University, Cleveland, OH  44106, USA} \affiliation{Department of Physics, Princeton University, Princeton, NJ 08544, USA}
\author{G.~Wang} \altaffiliation[Now at ]{Materials Science Division, Argonne National Laboratory, IL 60439, USA} \affiliation{Department of Physics, Case Western Reserve University, Cleveland, OH  44106, USA}
\author{S.~Yellin} \affiliation{Department of Physics, Stanford University, Stanford, CA 94305, USA} \affiliation{Department of Physics, University of California, Santa Barbara, CA 93106, USA}
\author{J.~Yoo} \affiliation{Fermi National Accelerator Laboratory, Batavia, IL 60510, USA} 
\author{B.A.~Young} \affiliation{Department of Physics, Santa Clara University, Santa Clara, CA 95053, USA} 

\collaboration{CDMS Collaboration} \noaffiliation

\date{\today}

\begin{abstract}

Data taken during the final shallow-site run of the first tower of the Cryogenic Dark Matter Search (CDMS~II) detectors have been reanalyzed with improved sensitivity to small energy depositions.  Four $\sim$224\,g germanium and two $\sim$105\,g silicon detectors were operated at the Stanford Underground Facility (SUF) between December 2001 and June 2002, yielding 118 live days of raw exposure.  Three of the germanium and both silicon detectors were analyzed with a new low-threshold technique, making it possible to lower the germanium and silicon analysis thresholds down to the actual trigger thresholds of $\sim$1\,keV and $\sim$2\,keV, respectively.  Limits on the spin-independent cross section for weakly interacting massive particles (WIMPs) to elastically scatter from nuclei based on these data exclude interesting parameter space for WIMPs with masses below 9\,GeV/$c^2$.  Under standard halo assumptions, these data partially exclude parameter space favored by interpretations of the DAMA/LIBRA and CoGeNT experiments' data as WIMP signals, and exclude new parameter space for WIMP masses between 3\,GeV/$c^2$ and 4\,GeV/$c^2$.

\end{abstract}

\pacs{95.35.+d, 14.80.Ly, 95.30.Cq, 29.40.Wk, 95.30.-k, 85.25.Oj}

\maketitle

\section{\label{sec:1}Introduction}

Astrophysical evidence strongly suggests that matter constitutes approximately one quarter of the energy density of the Universe.  Baryons in stars and intergalactic gas account for only a small fraction of the matter density, while the majority of the Universe's matter is of an unknown composition, collectively termed dark matter (DM) due to its apparently nonluminous nature~\cite{dm5}.  Observations of large-scale structure and supernovae, combined with measurements of the cosmic microwave background, imply a total matter and energy density close or equal to the critical density~\cite{wmap1}.  In terms of the critical density, the matter and energy budget breaks down as follows:
\begin{equation}
	\Omega_{\mathrm{baryon}} = 0.0456 \pm 0.0016,
	\label{eq:1}
\end{equation}
\begin{equation}
	\Omega_{\mathrm{DM}} = 0.227 \pm 0.014,
	\label{eq:2}
\end{equation}
and
\begin{equation}
	\Omega_{\Lambda} = 0.728^{+0.015}_{-0.016},
	\label{eq:3}
\end{equation}
where $\Omega_\Lambda$ represents the mysterious dark energy thought to be responsible for the current accelerating expansion of the Universe~\cite{wmapnew}.

The standard model of particle physics provides a single candidate for this nonbaryonic dark matter:  the neutrino.  Large-volume neutrino observatories have successfully measured and confirmed the existence of neutrino mass~\cite{superk}.  Nevertheless, neutrinos make only a small contribution to the dark matter density,
\begin{equation}
	\Omega_{\nu} = \frac{\sum m_{\nu}}{93.14\mathrm{\,eV} h^{2}} < 0.006,
	\label{eq:4}
\end{equation}
where $\sum m_{\nu} < 0.28$\,eV and $h$ is the dimensionless Hubble parameter~\cite{wmap2,*neutrino1,wmapnew}.  Furthermore, formation of large-scale structure in the Universe constrains the neutrino component of the dark matter~\cite{structure1,*structure2}.  Simulations of structure formation require a significant nonrelativistic, or ``cold,'' dark matter density~\cite{structure3}, which cannot arise from standard model neutrinos.

A number of suitable dark matter candidates arise from theories that propose physics beyond the standard model.  Weakly interacting massive particles (WIMPs)~\cite{dm2} are the most studied class of such dark matter particles.  In particular, many $R$-parity-conserving weak-scale supersymmetric (SUSY) theories offer a natural dark matter candidate in the form of the lightest superpartner~\cite{dm4,*dm5,dm1}, often a neutralino.  Massive, electrically neutral, and stable, the lightest neutralino, $\tilde{\chi}_{1}^{0}$, of many SUSY theories is an excellent WIMP candidate.  SUSY theories contain a vast space of unknown free parameters, which are constrained by requiring consistency with existing empirical particle physics and astrophysics knowledge.  Popular techniques for additionally restricting the extent of the free parameter space result in a lower bound on the $\tilde{\chi}_{1}^{0}$ mass of $\sim$40\,\gev~\cite{lep1,*lep2,*lep3}.  Efforts to explore a wide range of SUSY free parameter space indicate that $\tilde{\chi}_{1}^{0}$ masses as low as a few \gev can be accommodated~\cite{lsp1,*lsp2,*lsp3,*lsp4}.  Under some scenarios, a relatively light WIMP could resolve the apparent conflict between the DAMA/LIBRA annual modulation signal and the null results of other experiments~\cite{savage1,*savage2,*savage3}.

\begin{figure}[t!]
\centering
\includegraphics[width=3.4in]{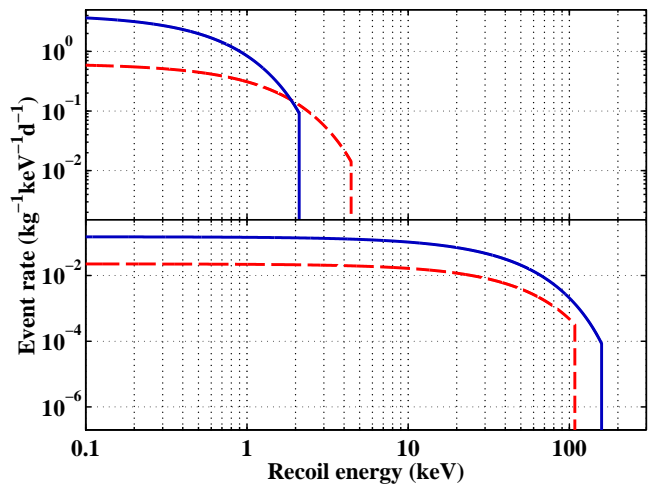}
\caption{(color online).  Expected differential event rates for 5\,\gev (top panel) and 100\,\gev (bottom panel) WIMPs scattering from Ge (blue/solid) and Si (red/dashed) targets.  All event rate calculations are based on the standard halo model described in \cite{halo1}, for an arbitrarily chosen WIMP-nucleon cross section of 1x10$^{-41}$\,cm$^{2}$.  Each energy spectrum  cuts off abruptly at a maximum recoil energy due to the assumed galactic escape velocity.  We use the 544\,km/s galactic escape velocity from \cite{rave1}, while all other halo parameters are taken from \cite{halo2}, and the local WIMP density is assumed to be 0.3\,GeV/cm$^{3}$.}
\label{fig:WIMPrates}
\end{figure}

If WIMPs are the dark matter, they form a spherical cloud (or halo) in which the luminous portions of the Milky Way are embedded, and will scatter very rarely off the nuclei in terrestrial matter.  Direct detection experiments seek to observe and measure the kinetic energy, in the keV range, of the recoiling nuclei.  The expected spectrum of WIMP-induced nuclear recoils decreases rapidly with increasing recoil energy, with the mean recoil energy directly proportional to the reduced mass of the WIMP-nucleus system.  Events with the lowest recoil energies are most numerous for all WIMP masses, and thus direct detection experiments generally strive for low recoil energy thresholds.  Sensitivity to low recoil energies is particularly crucial for experiments seeking to detect light WIMPs.  Figure~\ref{fig:WIMPrates} illustrates expected spectra of nuclear-recoil energies for two types of target nuclei for two WIMP masses.

The advantages of a low threshold must be weighed against the degradation of background rejection capability at low recoil energies.  Furthermore, the intrinsic rates of some categories of background may also increase at low energies.  Finally, special care is required to account for the effects of non-zero energy resolution at energies near the electronic-noise level.

In this work we describe a new analysis of data from the Cryogenic Dark Matter Search (CDMS) experiment, with special attention to events of low recoil energy.  In order to access this low-energy parameter space, we forgo the pulse-shape discrimination techniques used to reject near-surface background events in previous CDMS analyses.  In contrast to previous CDMS results, the signal region of this analysis will be populated by a number of background events.  We adopt an inclusive philosophy that maximizes the detection efficiency at low energy while limiting the rate of nonphysical sources of background, such as electronic noise.  This low-threshold analysis sacrifices some of the strengths of the CDMS experiment's traditional background discrimination methods for a chance to probe previously untested low-mass WIMP parameter space.  

\section{\label{sec:2}Experimental Setup and Technique}

\subsection{\label{sec:2a}Apparatus}

The data reported on here were recorded during the final exposure of the first six CDMS II detectors at the Stanford Underground Facility (SUF)~\cite{cdms1,cdms2}.  The SUF setup provided a high level of shielding against external sources of radiation.  The SUF was a shallow site with $\sim$17 meters water equivalent overburden, effectively stopping hadronic cosmic rays and reducing the muon flux by a factor of 5.  The remaining incident muons were tagged with a high-efficiency, hermetic plastic scintillator muon veto, allowing offline rejection of muon-coincident detector interactions.  The muon veto enclosed several layers of tightly packed passive shielding.  A 15\,cm-thick outer lead shield and 25\,cm-thick outer polyethylene shield surrounded the detector cold volume to attenuate external photons and degrade external neutrons, respectively.  Inside the radiopure copper walls that delineated the innermost 20\,mK cold volume, 1\,cm of ancient lead and an additional 11\,kg of polyethylene surrounded the detector assembly, providing further shielding.

\begin{table}[t!]
\renewcommand{\arraystretch}{1.3}
\caption{The first six CDMS II detectors are listed in order of their relative positions (\textit{from top to bottom}) within the detector tower, indicating each detector's name, material, and mass.}
\label{tab:det_masses}
	\begin{center}
		\begin{ruledtabular}
			\begin{tabular*}{3.4in}{@{\extracolsep{\fill}} @{\hspace{0.2in}}c c c@{\hspace{0.2in}}} 
				Name&Material&Mass (grams)\\   \hline 
				Z1&Ge&230.5\\
				Z2&Ge&227.6\\
				Z3&Ge&219.3\\
				Z4&Si&104.6\\
				Z5&Ge&219.3\\
				Z6&Si&104.6\\ 
			\end{tabular*}
		\end{ruledtabular}
	\end{center}
\end{table}

In the \ center of \ the apparatus six \ Z-sensitive Ionization- and Phonon-mediated (ZIP) detectors~\cite{zip1,*zip2,*zip3} were arranged in a vertical stack (``tower''), with adjacent detectors separated by 2.2\,mm with no intervening material.  Table~\ref{tab:det_masses} indicates the detector names, materials, masses and relative positions within the tower.

Each detector has two ionization electrodes deposited on its bottom surface; a circular inner electrode (``q-inner'') covering 85\% of the physical area, and an annular outer electrode (``q-outer'') which permits identification and rejection of events with energy depositions within the outer detector volume.  Photolithographed onto the top side of each Ge (Si) detector were 4144 (3552) Al and W superconducting Quasiparticle-trap-assisted Electrothermal-feedback Transition-edge sensors (QETs).  The 1036 (888) QETs in a given Ge (Si) detector quadrant are electrically connected, resulting in four individually read out phonon sensors whose shared borders orthogonally bisect the surface.  By measuring timing and pulse height differences between the sensors, we can reconstruct an event's position in the plane parallel to the detector's top and bottom surfaces (``xy-position'').

Following amplification by a SQUID array~\cite{squid1,*squid2} and room temperature electronics, two copies of each detector's four phonon signals were generated at the hardware level.  A band-pass filtered analog sum of one set of phonon signals (the ``triggering phonon energy'') was compared to a low-level discriminator threshold.  The resulting logical pulses were OR'ed across all six detectors to form the experimental trigger.  The second set of phonon signals were individually digitized, subjected to a software optimal filter, and summed to constitute the ``reconstructed phonon energy.''  The hardware band-pass filter's poles were chosen to resemble the software filtering as closely as possible.  The energy deposition used to trigger the data acquisition system is therefore very similar to the energies evaluated in software during offline analysis, but not exactly the same.  The discriminator thresholds were carefully tuned to the lowest levels possible to ensure that the overall trigger rate of $\lesssim$1\,Hz during WIMP search was dominated by true particle interactions, while simultaneously allowing occasional triggers due to electronic-noise fluctuations.

\subsection{\label{sec:2b}Measurement technique}

Energy deposited by recoils causes two types of signals in our detectors.  Most of the energy is deposited as a spectrum of high frequency athermal phonons.  In addition, electron-hole pairs are created.  The average deposited energy per pair created is $\epsilon$=3.0 (3.8)\,eV for Ge (Si)~\cite{fano1,*fano2} (see also Appendix~C in \cite{fano3} for a detailed discussion of $\epsilon$).  To measure the ionization, the QET side of our detectors was held at ground, and a bias voltage between -3\,V and -6\,V was applied to the electrodes on the other detector side, causing drift of liberated electrons (holes) to the QET (electrode) side.  Charge-sensitive amplifiers with field-effect transistor front-ends amplified the signals on the electrodes and enabled inference of the ``ionization energy'' $Q$ from the total liberated ionization.  The total phonon signal is comprised of three parts:  1)phonons produced promptly by the recoil (``primary phonons''); 2)phonons produced by the drifting ionization (``Neganov-Luke phonons'' or ``drift heat''~\cite{phonon1,*phonon2}); and 3)phonons produced when electrons and holes recombine, usually at the detector surfaces (``recombination phonons'').  The QETs and their amplification system enabled deduction of the energy $P_{\mathrm{total}}$ from the total phonons.  The ``recoil energy,'' $E_{\mathrm{recoil}}$, is equal to the sum of the primary and recombination phonon energies, and can be deduced by subtracting off the drift heat:
\begin{equation}
	E_{\mathrm{recoil}} = P_{\mathrm{total}} - \frac{eV}{\epsilon}Q,
	\label{eq:5}
\end{equation}
where $V$ is the absolute value of the bias voltage, and $e$ is the charge of the proton.  

The CDMS detectors use these two sensor technologies to discriminate nuclear recoils, produced by WIMP candidates, from much more numerous electron recoils, produced mostly by background photons.  The detectors provide a simultaneous measurement of ionization and phonons for each particle-interaction event within the target.  The ratio of ionization to recoil energy (``ionization yield'' $Y\equiv Q/E_{\mathrm{recoil}}$) is higher for electron recoils than for nuclear recoils, and provides near-perfect event-by-event discrimination for recoil energies in excess of 10\,keV.  However, the ionization yield is broadened by electronic noise for recoil energies lower than 10\,keV, and discrimination power is lost for recoil energies below about 2\,keV.

The energy scale for $Q$ and $E_{\mathrm{recoil}}$ was calibrated by electron recoils caused by gamma-ray sources.  The resulting ionization yield for electron recoils is (on average) equal to unity by construction.  However, nuclear recoils cause less ionization, and generally follow the theory by Lindhard \etal~\cite{lind1,*lind2}.  Nuclear recoils from a neutron source allowed determination of the energy dependence of their ionization yield, which smoothly increases from $\sim$0.2 at $E_{\mathrm{recoil}}$=2\,keV to $\sim$0.4 at 100\,keV.

The ionization yield for nuclear recoils with recoil energies less than 2\,keV is difficult to measure directly with ZIP detectors.  However, measurements using Ge detectors have been performed for energies as low as $\sim$0.25\,keV (see Fig.~3 in Appendix~III in \cite{texono} for example), and agree well with Lindhard \etal's model.  We therefore extrapolate via a power law from 2\,keV to lower energies such that the ionization yield is zero at 0\,keV.  The extrapolation is needed only to estimate the energy dependence of a few detection efficiencies for the lowest-energy nuclear recoils.  At these low energies, the recoil energy is less than (but nearly equal to) $P_{\mathrm{total}}$, and the extrapolation is used to estimate recoil energy from $P_{\mathrm{total}}$ via a small correction.  Any uncertainty related to this small correction is therefore small as well.  Alternative extrapolation methods have been used to test the effect that uncertainty in the ionization yield for low-energy nuclear recoils might have on our ability to detect light WIMPs.  We have determined that any systematic uncertainty introduced by our particular choice is small compared to the other sources of uncertainty discussed in Section~\ref{sec:4d}.

As discussed earlier, in this paper we extend the lowest recoil energies to the 1\,keV (2\,keV) range for Ge (Si) targets.  Nuclear recoils with such low energies will appear to deposit very small ionization energy, in the tenths of keV.  Although earlier CDMS analyses implemented minimum requirements on the ionization energy, in this analysis we have avoided any minimum requirement on $Q$.  We emphasize that our hardware trigger used only the total phonon energy, which slightly exceeds the recoil energy for nuclear recoils.

\subsection{\label{sec:2c}Recoil-energy estimators}

One way to estimate the recoil energy is with both measured $P_{\mathrm{total}}$ and measured $Q$ used in Eq.~\ref{eq:5} on an event-by-event basis.  This method incorporates errors on both these measured quantities into the estimate of $E_{\mathrm{recoil}}$, and introduces significant correlations between $E_{\mathrm{recoil}}$ and the ionization yield.  This is the traditional CDMS method for estimating recoil energy because it is accurate even if it is not known whether an event is an electron or nuclear recoil.  The analysis cuts and their efficiencies (described in Section~\ref{sec:3b}) use this event-by-event recoil-energy estimate when needed.  We refer to this estimate as the ``$Q$-corrected'' recoil energy to distinguish it from an alternative method discussed below.  Throughout this paper, the recoil energy is always $Q$-corrected unless otherwise stated.

The recoil energy can also be estimated from $P_{\mathrm{total}}$ alone by scaling it to reflect the average ionization yield response measured from calibration samples.  For example, since the ionization yield for electron recoils is on average equal to one, the ionization energy for electron recoils is on average equal to the recoil energy itself.  Replacing $Q$ by $E_{\mathrm{recoil}}$ in Eq.~\ref{eq:5} yields $E_{\mathrm{recoil}}$ equal to one-half (one-third) of $P_{\mathrm{total}}$ for Ge ZIP detectors operated with a -3\,V (-6\,V) bias voltage.  We refer to this estimate as the recoil energy corrected by electron-recoil ionization yield, or ``$Y_{\mathrm{ER}}$-corrected'' recoil energy.  Due to its superior spectral resolution, it is particularly useful when studying ZIP-detector response to x-ray and gamma-ray sources (see Fig.~\ref{fig:xrays} for example).

A similar recoil-energy estimate can be made for nuclear recoils.  The reduced drift heat associated with nuclear recoils is energy dependent, and can be subtracted according to the mean ionization yield measured from \cf calibrations (dashed line in Fig.~\ref{fig:cf_nrb} for example).  The resulting recoil-energy estimator has the advantage of including electronic noise from only the phonon channel.  We refer to this estimate as the recoil energy corrected by nuclear-recoil ionization yield, or ``$Y_{\mathrm{NR}}$-corrected'' recoil energy.  Discussed in more detail in Section~\ref{sec:3a}, the hardware and software energy thresholds employed in this analysis depend solely on the total phonon signal.  The associated threshold efficiencies (see Figs.~\ref{fig:TrigEff} and ~\ref{fig:threshEff}, and Table~\ref{tab:thresholds} for example) for detecting nuclear recoils are therefore functions of the $Y_{\mathrm{NR}}$-corrected recoil energy.

\subsection{\label{sec:2d}Data samples}

The first tower of CDMS II ZIPs was commissioned at the SUF in the second half of 2001, and WIMP-search data were recorded between December 2001 and June 2002.  Collectively termed ``Run~21,'' the WIMP search was split into two distinct operational periods.  The Ge (Si) detectors were initially operated with a -3\,V (-4\,V) bias voltage for 94 days between December 2001 and April 2002, yielding just over 66 live days (henceforth referred to as the ``3V data'').  These data were originally analyzed with a 5\,keV analysis threshold, and the resulting exclusion limit on the spin-independent WIMP-nucleon cross section is still one of the strongest constraints on low-mass WIMPs~\cite{cdms1}.  An additional 52 live days of previously unpublished data were recorded over a period of 74 days between April and June 2002.  In an attempt to reduce the surface-event background, we experimented with a -6\,V bias voltage during this latter part of the run (henceforth referred to as the ``6V data'').  Although the larger bias voltage improved charge collection for events near the detector surfaces, it also degraded the phonon pulse rise-time information used to reject surface events.  Overall, the rate of surface events leaking into the signal region following a surface-event rejection analysis was slightly greater for the 6V data.  CDMS ZIP detectors have since been operated with the lower bias voltage settings.  In this paper we analyze data from both charge bias runs.

A mixture of WIMP-search and calibration data was recorded during Run~21.  Detector response to electron recoils was tested by introducing a \co gamma-ray source just inside the outer lead shielding.  Three gamma-ray calibrations were performed for each voltage bias.  Relatively short exposures conducted before recording any WIMP-search data were followed by more extensive exposures midway through and after WIMP-search runs.  These data were primarily used for an event-by-event correction of the phonon signals for observed position- and energy-dependent response functions (``position correction'').  The xy-position information available from the modularity of the phonon sensors was used to correct variations in the reconstructed optimal-filter pulse height due to variation in pulse shape with event position, yielding improved phonon energy resolution.

With part of the outer polyethylene shielding removed, the detectors were also exposed to a \cf neutron source to calibrate their response to nuclear recoils.  Neutron calibrations were performed at the beginning and end of the WIMP searches for each bias voltage.  The ionization yield distributions obtained from these data helped define the nuclear-recoil signal region that WIMPs are expected to populate.  Nuclear-recoil detection efficiencies were estimated from these data as well.

Although the hardware thresholds were tuned to occasionally trigger on electronic noise, the rate was far too low for proper characterization of the distribution of subthreshold noise pulses, and for estimates of near-threshold efficiencies.  To understand the near-threshold phenomena, samples of detector performance without the trigger bias (``nontriggered data'') are critically important.  During Run~21, nontriggered data were obtained by invoking a randomly generated software trigger, resulting in a periodic sampling of the electronic-noise environment.  These events constitute $\sim$5\% of all triggers, and can be considered a third type of calibration data.  Due to the low rate of true particle interactions in the detectors, randomly triggered events usually consisted of only noise fluctuations, with a very low probability of measuring a non-zero energy in any of the detectors' sensors.  While useful for determining daily noise levels, the randomly triggered data are inadequate for efficiency estimates since so few have energies near the detector thresholds.  Furthermore, due to a timing artifact associated with the implementation of the random trigger, these data were plagued by a low rate of unrepresentative noise traces.  When reconstructed, the problematic traces contribute non-Gaussian tails to each detector's underlying electronic-noise distribution.  Consequently, if the randomly triggered noise distributions are scaled to an exposure equivalent to the WIMP-search data and then subjected to the WIMP-search cuts, the resulting rate of noise pulses in the signal region is grossly overestimated.  We therefore made no use of nontriggered data for characterizing resolutions and efficiencies.

Fortunately, an alternate sample of data that avoids the trigger bias was acquired during normal, triggered readout.  Although most triggers were associated with an energy deposition in a single detector, traces for all six detectors were recorded and analyzed.  Data from the detectors that did not trigger (``other-detector triggers'') provide a fair and representative sample of each detector's subthreshold noise distribution.  Provided the noise distribution is constructed from events where the triggering detector was not an adjacent detector, non-Gaussian tails are avoided.  Events for which the triggering detector was either directly above or below often contained small energy depositions due to true multiple-detector interactions.  This class of events includes a sampling of recoil energies up to and exceeding the detector trigger thresholds, providing events with which to probe near-threshold behavior.  Some of these had energies exceeding their detector's trigger threshold and are not truly nontriggered events, but rather events with a delayed trigger that occurred during a readout instigated by another detector.  A logical-pulse-based post-trigger history of when these delayed triggers occurred was recorded for each detector and event, and is the basis of the hardware trigger efficiency estimates detailed in the next section.

\section{\label{sec:3}Data Analysis}

\subsection{\label{sec:3a}Noise and thresholds}

\begin{figure}[t!]
\centering
\includegraphics[width=3.4in]{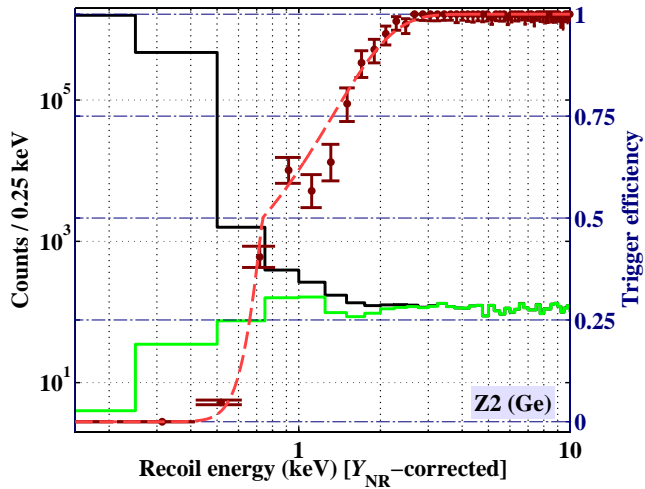}
\caption{(color online).  Hardware trigger efficiency estimate (error bars) for a representative Ge detector (Z2 3V data) as a function of $Y_{\mathrm{NR}}$-corrected recoil energy.  The efficiency is calculated in bins of 0.25\,keV by dividing the distribution of energies correlated to the presence of logical pulses in the post-trigger history (green/light solid) by the distribution of all reconstructed energies (black/dark solid).  The efficiency scale is given by the y-axis on the right (blue/dash-dotted grid lines), while the scale for the histograms is given by the y-axis on the left (black/dotted grid lines).  A split-width error function (red/dashed) fits the efficiency estimate over the full range of energies, yielding a hardware threshold of 0.74\,keV at 50\% efficiency.  For this detector the hardware trigger is 100\% efficient for energies above $\sim$3\,keV.}
\label{fig:TrigEff}
\end{figure}

To be considered viable WIMP candidates, events were required to exceed hardware and software thresholds in phonon energy.  We used data from other-detector triggers to measure the corresponding detection efficiencies as a function of reconstructed phonon energy.  Since the triggering phonon energy was slightly different from its offline reconstructed counterpart, the hardware threshold efficiencies as a function of the latter are not simple step functions.  To characterize the hardware threshold efficiencies, we first evaluate the reconstructed phonon energy in the 50\,$\mu$s following the time of each other-detector trigger.  Generally these reconstructed phonon energies are Gaussian distributions centered at zero energy, consistent with electronic noise (``noise cores'').  Occasionally nonzero energies were reconstructed, augmenting each noise core with an approximately uniform distribution extending to higher energies.  Sometimes the presence of a logical pulse from the low-level discriminator threshold was recorded in the post-trigger history of an other-detector trigger.  We used events with logical-pulse presence to obtain the distribution in reconstructed phonon energy of successful triggers, and then divided by the distribution of all (unsuccessful and successful) triggers to characterize the hardware threshold efficiency, as illustrated for one Ge detector in Fig.~\ref{fig:TrigEff}.

\begin{table}[t!]
\renewcommand{\arraystretch}{1.6}
\caption{The phonon energy thresholds (at 50\% efficiency) with 1$\sigma$ errors (68.3\%\,C.L.) for the viable low-threshold detectors in terms of $Y_{\mathrm{NR}}$-corrected recoil energy.  For each detector the hardware trigger threshold and the average 6$\sigma$ noise rejection software threshold is listed for its 3V and 6V data.}
\label{tab:thresholds}
	\begin{center}
		\begin{ruledtabular}
			\begin{tabular*}{3.4in}{@{\extracolsep{\fill}} l c c c c} 
				& \multicolumn{2}{c}{Hardware (keV)} & \multicolumn{2}{c}{Software (keV)}\\ 
				Detector&3V&6V&3V&6V\\ \hline 
				Z2&$0.74^{+0.07}_{-0.02}$&$0.67^{+0.09}_{-0.02}$&$0.63\pm0.01$&$0.58\pm0.02$\\
				Z3&$1.13^{+0.07}_{-0.04}$&$1.12^{+0.07}_{-0.05}$&$0.82\pm0.02$&$0.72\pm0.03$\\
				Z4&$1.77^{+0.12}_{-0.10}$&$1.71^{+0.13}_{-0.10}$&$1.62^{+0.10}_{-0.09}$&$1.52^{+0.09}_{-0.08}$\\
				Z5&$1.00^{+0.05}_{-0.04}$&$0.91^{+0.06}_{-0.05}$&$0.73\pm0.02$&$0.63^{+0.04}_{-0.03}$\\
				Z6&$1.53^{+0.08}_{-0.06}$&$1.55^{+0.09}_{-0.06}$&$1.39^{+0.06}_{-0.05}$&$1.34\pm0.05$\\ 
			\end{tabular*}
		\end{ruledtabular}
	\end{center}
\end{table}

A reconstructed phonon energy threshold (``software threshold'') was applied based upon the essentially Gaussian behavior of the reconstructed phonon energy.  A Gaussian fit to the noise core of each day's randomly triggered events was performed, and we required that events exceed a threshold of 6$\sigma$ above the mean (for that day) in reconstructed phonon energy to be considered a WIMP candidate.  The 6$\sigma$ thresholds for the Z2 and Z3 3V data were the most stable, varying by less than 5\% from day to day, while those for the Z1 data were the least stable, varying by as much as 50\%.  The software threshold efficiencies were calculated using the same method employed for the hardware trigger efficiency estimates described above, and differ from step functions because of time variations in the Gaussian fit parameters.  Table~\ref{tab:thresholds} summarizes the average hardware and software phonon energy thresholds.

The topmost Ge detector (Z1) is not included in Table~\ref{tab:thresholds} because it was rejected as a low-threshold detector.  Z1 suffered from particularly strong dependence of pulse height on xy-position for which the gamma calibration position correction was unable to fully compensate.  The corrected energy resolution and associated 6$\sigma$ thresholds are three to four times larger than those of the other Ge detectors, yielding not only a larger analysis threshold, but also causing the ionization yield-based discrimination to break down at a higher energy.  Since a detector's low-mass WIMP sensitivity is critically dependent on its detection threshold, data from Z1 would contribute very little to the reach of this analysis while allowing a disproportionate number of background events to leak into the signal region.  Z1 was only used to veto multiply scattering events.

\begin{figure}[t!]
\centering
\includegraphics[width=3.4in]{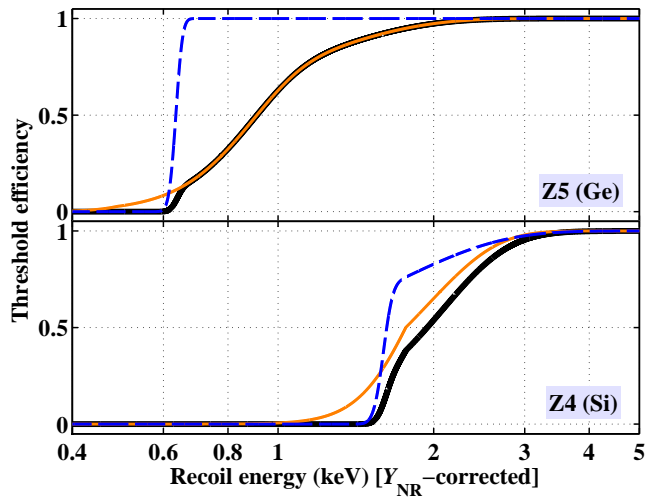}
\caption{(color online).  Fits to binned estimates of the hardware (orange/light solid) and software (blue/dashed) phonon energy threshold efficiencies are multiplied, yielding the combined threshold efficiencies (black/dark solid) for typical Ge (top panel, Z5 6V data) and Si (bottom panel, Z4 3V data) detectors as a function of $Y_{\mathrm{NR}}$-corrected recoil energy.}
\label{fig:threshEff}
\end{figure}

Each detector's final threshold efficiency is a product of its hardware and software threshold efficiencies.  As a function of reconstructed phonon energy, the software threshold efficiencies nearly resemble step functions, rising to 100\% more quickly and at lower energies than the hardware efficiencies.  Consequently, the combined efficiencies differ from the hardware trigger efficiencies only for very low energies.  Figure~\ref{fig:threshEff} illustrates the combined efficiency for typical Ge and Si detectors.  For all but one detector the combined thresholds at 50\% efficiency are equal to the hardware thresholds listed in Table~\ref{tab:thresholds}.  During one week of the 3V data run, Z4's phonon noise was abnormally high, causing an extended period of higher software thresholds.  The effect on the software and combined efficiencies is shown in Fig.~\ref{fig:threshEff}.  The resulting combined threshold of 1.9$\pm$0.1\,keV is the largest among the accepted low-threshold detectors.

\subsection{\label{sec:3b}Analysis cuts and efficiencies}

WIMP candidates were required to pass a variety of stringent data-quality cuts.  Events for which any part of the data record was incomplete or inconsistent were excluded.  Excessively noisy traces as well as traces with multiple pulses (``pileup'') were rejected based on the performance of the optimal-filter fits to the ionization signals.  Fits with unusually large $\chi^2$ values were removed from consideration.  Noisy traces and pileup were further suppressed by requiring the variances of the phonon and ionization traces' prepulse baselines to be within 5$\sigma$ of the average behavior for randomly triggered noise traces.  Together, the data-quality cuts exhibit no energy dependence, and have a combined WIMP detection efficiency of $\sim$99\%.  This applies for each detector type and for both the 3V and 6V data.

If this analysis were restricted to recoil energies greater than 2\,keV, the above data-quality cuts would be sufficient.  Throughout the 6V data run, however, the second Ge detector (Z2) was subject to intermittent periods of high trigger rates (``event bursts'') due to sub-2\,keV pulses.  Many followed cryogenic and detector servicing periods during which detector temperatures were elevated.  Consequently, we conclude that these event bursts were not due to physical recoils in the detector, and were considered periods of poor data quality.  Close examination of the phonon traces for these events revealed elevated prepulse baselines with nominal variability, allowing them to pass the standard data-quality cut described above.  Discrimination parameters based on the traces' average prepulse baselines were developed to cut the event bursts.  Application of the cut reduces Z2's 6V WIMP search from 52 days to 20 days, while reducing the corresponding number of WIMP candidates by a factor of $\sim$30.  Despite being isolated to a single detector and less than half of the WIMP-search data, without this cut the burst events would be the dominant source of background invading the signal region.

A fiducial-volume cut based on the ionization signals rejects recoils that occurred near detector edges.  The q-inner signal estimates an event's ionization energy, while the q-outer signal is required to be consistent with noise.  Crosstalk between the inner and outer electrodes adds a small contribution to the q-outer signal proportional to the q-inner pulse amplitude.  To model this dependence, the q-outer noise levels were parametrized as a function of q-inner energy, yielding 2$\sigma$ q-outer ``noise bands'' designed to accept 95.5\% of events within the fiducial volume.  Phonon sensors of the adjacent detector also induced signals in the electrodes, initially causing a loss of otherwise viable events.  To retain these events, the fiducial-volume cut was modified to reject only events that exceed their noise band's upper limit, increasing the cut's expected acceptance to 97.7\%.  Combining the acceptance with the physical coverage of the q-inner electrode yields an expected fiducial volume of $\sim$83\% of the gross detector mass.  We measured the efficiency of the fiducial-volume cut for recoil energies between 4\,keV and 100\,keV with nuclear-recoil calibration data from \cf exposures.  The average efficiency agrees well with physical expectation, varying between $\sim$81\% (Z4 3V data) and $\sim$83\% (Z2 3V data).  Over the measurable energy range, the efficiency of the cut exhibits a weak energy dependence, generally decreasing with increasing recoil energy.  The Z4 6V data are the most extreme, where the efficiency smoothly decreases from $\sim$84\% at 4\,keV to $\sim$79\% at 100\,keV.  For recoil energies less than 4\,keV, a nuclear-recoil event's q-inner and q-outer signals are difficult to distinguish from electronic noise, and the fiducial-volume cut is unable to differentiate events within the fiducial volume from those that occurred near a detector's outer edge.  Due to increased acceptance, the efficiency of the fiducial-volume cut should therefore increase rapidly as the recoil energy decreases toward 0\,keV.  Since we were unable to measure this low-energy behavior, we made the conservative choice to linearly extrapolate the efficiency for recoil energies less than 4\,keV to match the efficiency measured at 4\,keV. 

The energy deposited by a WIMP in one of our detectors would be so localized and infrequent that only events caused by backgrounds will cause significant energy depositions in two or more detectors simultaneously.  We therefore impose a ``single-scatter'' criterion, requiring signal events to have had a significant energy deposition in no more than one detector.  Since the phonon signal provides the most sensitive indicator for a particle interaction in our detectors, we based the single-scatter cut on the 6$\sigma$ software thresholds described above and in Table~\ref{tab:thresholds}.  An event is considered a single scatter if only one detector has a reconstructed phonon energy exceeding its software threshold.  With an experimental trigger rate of $\lesssim$1\,Hz, the probability of more than one pulse occurring within our $\lesssim$2\,ms digitization time is negligible.  The efficiency of the single-scatter cut is therefore nearly 100\%, with the near-threshold behavior taken into account by the software threshold efficiency described above and in Fig.~\ref{fig:threshEff}.

\begin{figure}[t!]
\centering
\includegraphics[width=3.4in]{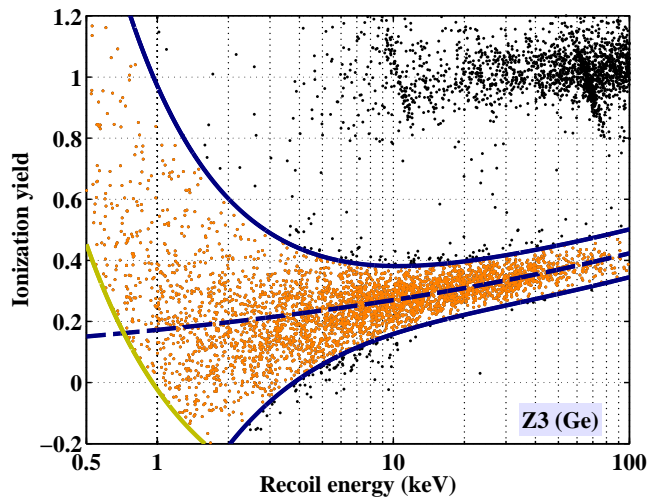}
\caption{(color online).  Ionization yield plotted as a function of recoil energy for representative Ge (Z3 6V data) \cf neutron calibration data, with the average ionization yield (blue/dashed), 2$\sigma$ nuclear-recoil band (blue/dark solid) and software phonon threshold (yellow/light solid) overlaid.  Unvetoed single scatters passing the fiducial-volume and data-quality cuts are displayed for events consistent with the nuclear-recoil criterion (orange/light dots), and some that are not (black/dark dots).  A substantial gamma-ray flux from the \cf source populates the ``electron-recoil band'' near an ionization yield of 1.  Two lines at $\sim$10.4\,keV and $\sim$66.7\,keV resulting from decays of Ge isotopes can be distinguished among the electron-recoil events.  As described in Section~\ref{sec:3c}, the low number of events above the nuclear-recoil band between 2\,keV and 6\,keV recoil energy qualitatively demonstrates that channeling does not significantly diminish this experiment's efficiency for detecting low-mass WIMPs.}
\label{fig:cf_nrb}
\end{figure}

Candidate events must occur when there is no activity in the muon veto.  A time history of muon veto activity (``veto hits'') was recorded for each ZIP-triggered event at an average rate of $\sim$5\,kHz.  This rate far exceeds the true rate of incident muons.  Due to large and awkward counter geometries, many of the plastic scintillator counters had regions with poor light collection.  To ensure efficient tagging of muons passing through these regions, the photomultiplier tubes monitoring each counter were operated at very high gains, causing regions with superior light collection (particularly near the photomultiplier tubes) to be partially sensitive to environmental gamma radiation.  We achieved a muon tagging efficiency of $>$99.9\% by rejecting a relatively high rate of ZIP detector interactions that were accidentally coincident with gamma rays registering as veto hits.  An event is vetoed if there were any veto hits in the 50\,$\mu$s to 80\,$\mu$s preceding the triggering phonon pulse, allowing for a significant difference in veto and phonon signal arrival times.  The veto signals were effectively instantaneous relative to the more slowly rising phonon pulses.  The longer delay was assigned to events with recoil energies $\lesssim$3.5\,keV, allowing smaller phonon pulses sufficient time to rise past the hardware trigger thresholds.  WIMP detection efficiency is lost due to veto hits from gamma-ray activity.  The detection efficiency that remains for each detector following application of the veto cut ranges from $\sim$67\% to $\sim$78\%.

\begin{figure}[t!]
\centering
\includegraphics[width=3.4in]{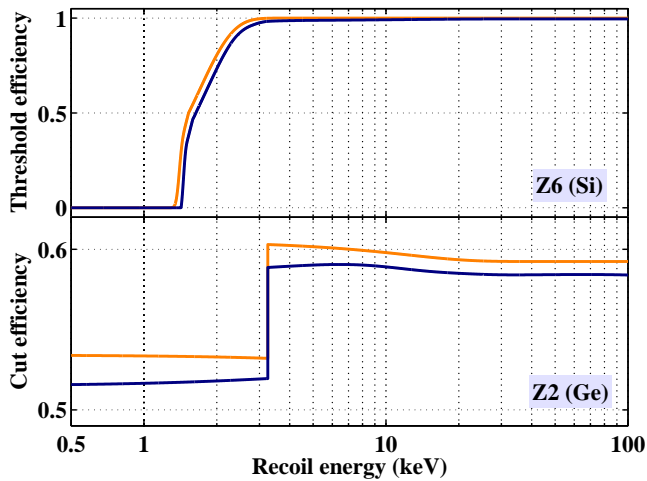}
\caption{(color online).  The detection efficiencies associated with the hardware and software phonon thresholds (top panel, Si Z6 3V data) and the combined analysis cuts (bottom panel, Ge Z2 6V data) for representative detectors.  The best-fit mean efficiency (orange/light solid) is compared in each case to its 90\% statistical lower limit (blue/dark solid).  The combined threshold efficiencies are plotted versus $Y_{\mathrm{NR}}$-corrected recoil energy, while the combined efficiency of the analysis cuts is plotted as a function of $Q$-corrected recoil energy.  The step in the latter is due to the larger muon-veto window chosen for the low-energy events.  Due to the uncertainty introduced by the efficiencies' statistical errors, we conservatively used the 90\% lower limit (1.28$\sigma$ below the mean) when calculating upper limits on the WIMP-nucleon cross section.}
\label{fig:reduced_eff}
\end{figure}

The final selection criterion requires a candidate event to have an ionization yield consistent with being a nuclear recoil, thereby discriminating against the otherwise overwhelming rate of electron recoils.  Several short exposures to the fission neutrons from a \cf source provided sufficient nuclear-recoil events to parametrize each detector's ionization yield response as a function of recoil energy.  When binned according to recoil energy, a detector's ionization yield distribution for nuclear recoils is well described by a Gaussian for energies as low as 2\,keV.  The mean ionization yield and associated 1$\sigma$ width as a function of recoil energy were estimated in a binwise fashion with maximum likelihood fits.  A simplified form of Lindhard \etal's theory for nuclear-recoil ionization yield in semiconductor crystals ($Y=a*E_{\mathrm{recoil}}^{b}$, where $a$ and $b$ are the fit parameters) was fit to the resulting collection of means, while an inverse-squared form was fit to the squared widths ($\sigma_{Y}^{2}=d^{2}+c^{2}/E_{\mathrm{recoil}}^2$, where $c$ and $d$ are the fit parameters).  The fitted ionization yield trends were used to construct nuclear-recoil acceptance bands, where an event is considered a nuclear recoil if its ionization yield lies within 2$\sigma$ of the mean.  An example nuclear-recoil band and mean ionization yield fit are compared to the \cf events from which they were formed in Fig.~\ref{fig:cf_nrb}.  The selection efficiency for each detector's band is well fit by an energy independent efficiency ranging from $\sim$93\% (Z6 3V data) to $\sim$96\% (Z3 and Z4 3V data).  

The combined efficiency of the analysis cuts as a function of recoil energy for a representative detector is shown in Fig.~\ref{fig:reduced_eff}.  The step between 3\,keV and 4\,keV is due to the larger muon-veto window chosen for the low-energy events.  The 90\% confidence level (statistical) lower limit efficiency (1.28$\sigma$ below the mean) is also shown, providing an indication of the statistical accuracy of the efficiency estimate for the analysis cuts.

\subsection{\label{sec:3c}Energy scale and resolution}

Our ZIP detectors are characterized by two energy scales:  recoil energy for electron recoils, and recoil energy for nuclear recoils.  The electron-recoil energy scale was initially calibrated with 662\,keV gamma rays from a \cs source as part of a series of short diagnostic and debugging runs conducted at the very beginning of Run~21.  The gamma rays emitted by \cs are lower in energy relative to the $\gtrsim$1\,MeV gamma rays emitted by \co, allowing easier calibration of the Si detectors for which such energetic photons are only partially contained.  

\begin{figure}[t!]
\centering
\includegraphics[width=3.4in]{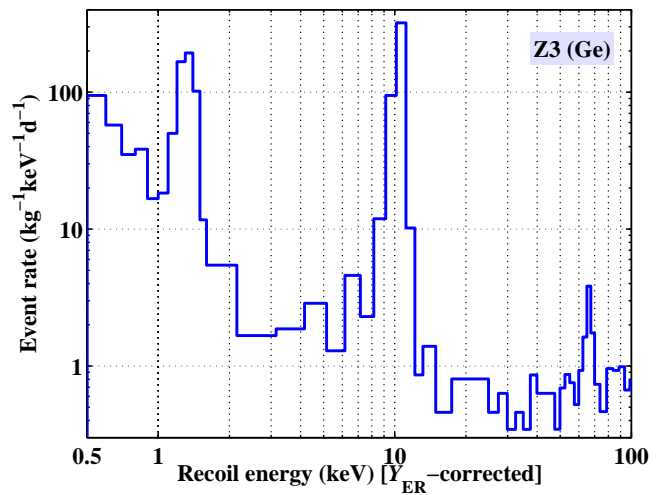}
\caption{Event rate of electron recoils for representative Ge (Z3 6V) WIMP-search data, plotted versus $Y_{\mathrm{ER}}$-corrected recoil energy.   Included events are unvetoed single scatters passing the data-quality and fiducial-volume cuts.  Three peaks due to internal decays of unstable Ge isotopes are clearly visible at 1.3\,keV, 10.4\,keV and 66.7\,keV.}
\label{fig:xrays}
\end{figure}

Each Ge detector's electron-recoil energy scale was confirmed and monitored using three distinct energy peaks that conveniently span our analysis energy range.  Both $^{68}$Ge and $^{71}$Ge are unstable isotopes; $^{68}$Ge is produced by cosmic rays, while $^{71}$Ge by thermal neutron capture.  The former is long lived with a half-life of $\sim$271 days and was primarily produced during detector construction and testing at sea level, while the latter is short lived with a half-life of 11.43 days and was primarily produced during exposure to the \cf source~\cite{isotopes}.  Both isotopes typically decay through electron capture followed by emission of x-rays or Auger-electrons, of 10.4\,keV for K-capture ($\sim$90\%) or 1.3\,keV  for L-capture ($\sim$10\%).  We measured the ratio of L- to K-captures to be 0.122$\pm$0.009, in good agreement with the ratio of 0.12 listed in \cite{isotopes}.  The cosmogenically-induced meta-stable $^{73m}$Ge state decays through successive emission of 53.4\,keV and 13.3\,keV photons, resulting in a 66.7\,keV peak at the upper end of our energy range.  The spectrum for an example Ge detector shown in Fig.~\ref{fig:xrays} clearly shows the three peaks from electron recoils induced by the above processes.  To improve spectral resolution, Fig.~\ref{fig:xrays} is plotted as a function of $Y_{\mathrm{ER}}$-corrected recoil energy.  A similar spectrum with corresponding features can be constructed from the ionization signal as well.

The Ge lines allowed measurement of the Ge phonon and ionization energy resolutions across our full analysis energy range.  The fractional energy resolutions for electron recoils vary from 8.8\% (Z5 6V phonon channel) to 26.0\% (Z3 3V ionization channel) at 1\,keV, and from 2.0\% (Z3 and Z5 3V ionization channels) to 5.4\% (Z2 3V phonon channel) at 100\,keV.  As there are no peaked features due to internal Si isotope decays, the same could not be done for the Si detectors except at zero energy, where resolution was determined solely by electronic noise.  

A low-threshold analysis must take into account the effect of non-zero energy resolution near the detection threshold.  The dominant contribution to the resolution near threshold for all detectors was from electronic noise.  We therefore dropped the energy-dependent terms of each detector's energy resolution when calculating WIMP exclusion limits, and simply applied the zero-energy resolution where needed.  Exclusion limits calculated with omission of the energy-dependent terms give more conservative limits than would be obtained with inclusion of those terms, and omission allows the Si and Ge detectors to be treated in the same manner.

As will be discussed in more detail in Section~\ref{sec:4c}, we made use of two forms of zero-energy resolution when calculating WIMP exclusion limits:  the $Y_{\mathrm{NR}}$-corrected and the $Q$-corrected recoil-energy resolutions.  The former is needed to include the effect of subthreshold pulses occasionally triggering the experiment due to phonon-noise fluctuations, and therefore does not include any noise contributions from the ionization channel.  The latter represents the resolution intrinsic to the combined efficiency of the analysis cuts, and includes ionization as well as phonon noise.  The other-detector triggered data were used to measure the $Y_{\mathrm{NR}}$-corrected and $Q$-corrected recoil-energy resolutions at 0\,keV.  The results are listed in Table~\ref{tab:resolution} for the viable low-threshold detectors.

\begin{table}[t!]
\renewcommand{\arraystretch}{1.4}
\caption{The 1$\sigma$ recoil-energy resolution at zero energy for the viable low-threshold detectors is listed for both the 3V and 6V data.  The $Y_{\mathrm{NR}}$-corrected resolution is based on the total phonon signal scaled to an equivalent recoil energy for nuclear recoils, while the $Q$-corrected resolution includes electronic-noise contributions from both the ionization and phonon signals.  All resolutions are rounded to the nearest eV and have an accuracy of $\sim$1\%.}
\label{tab:resolution}
	\begin{center}
		\begin{ruledtabular}
			\begin{tabular*}{3.4in}{@{\extracolsep{\fill}} l c c c c} 
				& \multicolumn{2}{c}{$Y_{\mathrm{NR}}$-corrected (eV)} & \multicolumn{2}{c}{$Q$-corrected (eV)}\\  
				Detector&{\hspace{.15in}}3V&{\hspace{.1in}}6V&{\hspace{.15in}}3V&{\hspace{.05in}}6V\\ \hline 
				Z2&{\hspace{.15in}}95&{\hspace{.1in}}88&{\hspace{.15in}}223&{\hspace{.05in}}387\\
				Z3&{\hspace{.15in}}126&{\hspace{.1in}}114&{\hspace{.15in}}282&{\hspace{.05in}}477\\
				Z4&{\hspace{.15in}}208&{\hspace{.1in}}196&{\hspace{.15in}}398&{\hspace{.05in}}538\\
				Z5&{\hspace{.15in}}113&{\hspace{.1in}}102&{\hspace{.15in}}258&{\hspace{.05in}}450\\
				Z6&{\hspace{.15in}}185&{\hspace{.1in}}179&{\hspace{.15in}}434&{\hspace{.05in}}678\\ 
			\end{tabular*}
		\end{ruledtabular}
	\end{center}
\end{table}

\begin{figure}[t!]
\centering
\includegraphics[width=3.4in]{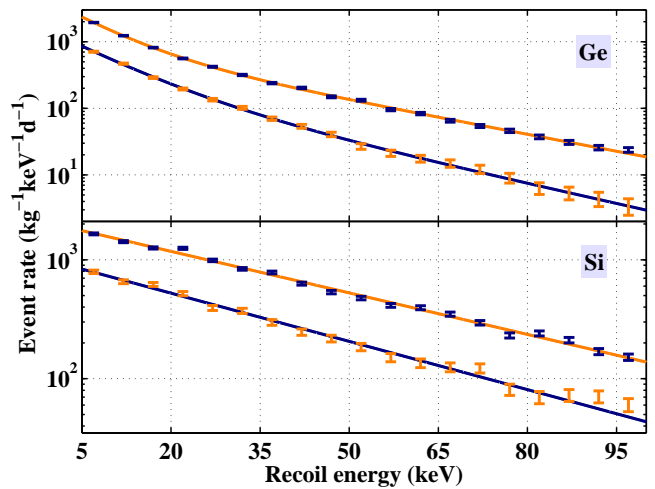}
\caption{(color online).  Comparison of Monte Carlo simulated recoil-energy spectra (blue/dark error bars) to experimentally measured recoil-energy spectra (orange/light error bars) from \cf calibrations for the Ge (top panel) and Si (bottom panel) detector ensembles.  Exponential fits are overlaid for both data (blue/dark solid) and Monte Carlo (orange/light solid).  Despite the clear disagreement in absolute event rate, comparison of the fitted decay constants provides a calibration of the nuclear-recoil energy scale for each target material.  There is a 16$\pm$3\% discrepancy in the Si energy scale.  A smaller discrepancy is observed in the Ge energy scale, and is within two standard deviations (statistical) of expectation.}
\label{fig:nr_escale2}
\end{figure}

Though the electron-recoil calibration is valuable for measuring detector performance, the nuclear-recoil energy scale is most important to WIMP searches.  We characterized this response by comparing the spectrum of nuclear recoils observed in \cf calibrations with a {\scshape geant3}~\cite{geant} simulation of the source and detector geometry.  In order to improve statistical power, we summed the observed events to produce mean Ge and Si recoil spectra.  Figure~\ref{fig:nr_escale2} shows the Monte Carlo and data distributions for both detector types with exponential fits, where the Si (Ge) spectra are well described by one (two) decaying exponential(s).  Comparison of the fitted decay constants tests our nuclear-recoil energy scale without being sensitive to the obvious disagreement in absolute rate.  Although difficult to perceive in Fig.~\ref{fig:nr_escale2}, there is a substantial 16$\pm$3\% discrepancy in the Si detectors' energy scale.  Without distinctive energy peaks with which to confirm the energy scale, we resolved the discrepancy by adjusting the Si recoil-energy scale upwards by 16\%, which has the effect of making the Si upper limit (discussed in Section~\ref{sec:4c}) more conservative.  The Si hardware and software thresholds and analysis cut efficiencies discussed above reflect this correction.  A smaller discrepancy exists for the Ge detectors, where the low-energy portion of the recoil distribution is overestimated relative to the Monte Carlo, and the high-energy portion is underestimated.  We did not make an analogous correction because the discrepancy is smaller (within two standard deviations of expectation), and the calibration done with electron-recoil peaks is more reliable.  The choice to not correct the Ge recoil-energy scale leads to a slightly more conservative exclusion limit for low-mass WIMPs.

The effect ``ion channeling'' might have on a direct detection experiment's energy scale has received considerable attention within the dark matter community since Sekiya \etal's~\cite{channel3} 2003 publication.  Early models~\cite{channel4,dama1} for a NaI scintillation detector suggested that a potentially significant fraction of recoiling nuclei might be ``channeled'' between rows and planes within a detector's crystalline lattice, allowing them to interact more weakly with atomic cores and deposit more energy in the form of ionization.  The full recoil energy from a WIMP interaction might therefore be detected via electron recoils, beneficially lowering a NaI detector's effective energy threshold for nuclear recoils. In our detectors, the larger ionization signal associated with channeling would cause an elevated ionization yield for nuclear recoils, tending to push channeled nuclear recoils closer to the electron-recoil population.  Ion channeling would therefore have a negative effect upon our nuclear-recoil acceptance efficiency.  If the effect is particularly pronounced for low recoil energies as suggested in \cite{dama1}, the low-mass WIMP sensitivity of this analysis could be severely impacted.

\begin{table}[t!]
\renewcommand{\arraystretch}{1.4}
\caption{The number of raw live days and corresponding number of WIMP candidate events for each detector and WIMP search following application of the analysis cuts and energy thresholds is listed.  The smaller exposure and corresponding number of candidates for the Z2 6V data is due to the event burst cut.}
\label{tab:candidates}
	\begin{center}
		\begin{ruledtabular}
			\begin{tabular*}{3.4in}{@{\extracolsep{\fill}} l c c c c} 
				& \multicolumn{2}{c}{3V Exposure} & \multicolumn{2}{c}{6V Exposure}\\ 
				Detector&Live Days&Candidates&Live Days&Candidates\\ \hline 
				Z2&66.12&159&20.16&67\\
				Z3&66.12&129&51.66&349\\
				Z4&66.12&130&51.66&125\\
				Z5&66.12&174&51.66&202\\
				Z6&66.12&401&51.66&314\\ 
			\end{tabular*}
		\end{ruledtabular}
	\end{center}
\end{table}

An upper limit on the fraction of channeled events near threshold can be calculated with the \cf data plotted in Fig.~\ref{fig:cf_nrb}.  For recoil energies between 2\,keV and 6\,keV, the fraction of events with ionization yield above the nuclear-recoil band is less than 4\%.  Although this rate is higher than the rate of electron recoils observed without the source present, neutron interactions with materials near the detectors can lead to secondary gamma rays.  In particular, neutron captures on hydrogen in the inner polyethylene shielding yield a continuum of electron-recoil energies due to the 2.2\,MeV photons released in the process, biasing our channeling upper limit high.  There were also several $\sim$1\,MeV photons emitted directly from the \cf source with each fission.  A limit on the channeling fraction that is less conservative, by incorporating a Monte Carlo  estimate of the rate of nonchanneled electron recoils during \cf calibrations, is beyond the scope of this paper.  Nevertheless, we do not see significant evidence of a channeling effect in our data that is large enough to appreciably affect the efficiencies estimated in this analysis.  We therefore ignore the effect of ion channeling, a decision that is supported by the recent and more sophisticated models developed by Bozorgnia \etal~\cite{channel1,*channel2}, which indicate that ion channeling for cryogenic Ge and Si targets is effectively nonexistent for low recoil energies.  

\section{\label{sec:4}Results}

\subsection{\label{sec:4a}Candidate events}

\begin{figure}[t!]
\centering
\includegraphics[width=3.4in]{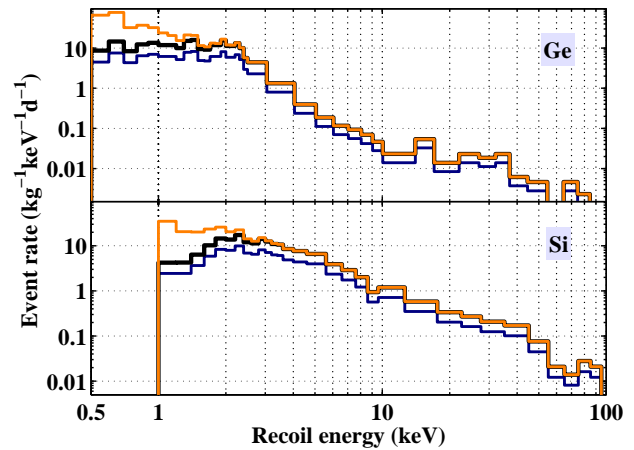}
\caption{(color online).  The combined Ge (top panel) and Si (bottom panel) WIMP candidate event rates as a function of recoil energy.  The uncorrected event rates (blue/dark thin solid) are compared to the efficiency corrected event rates.  The latter are successively corrected by the exposure-weighted detector-averaged efficiencies of the analysis cuts (black/dark thick solid), and then by the detector-averaged hardware and software thresholds (orange/light thin solid).}
\label{fig:nr_spectra}
\end{figure}

Following application of the analysis cuts and phonon software thresholds, a substantial residual rate of events is observed in the low-threshold signal region.  We restrict our attention to events with recoil energies between 0.5\,keV (1\,keV) and 100\,keV for the Ge (Si) detectors, yielding a total of 1080 Ge and 970 Si candidate events.  The number of raw live days and corresponding number of candidate events for each detector and bias voltage are listed in Table~\ref{tab:candidates}.  The combined Ge and Si recoil-energy spectra are shown in Fig.~\ref{fig:nr_spectra}, where the event rates have been successively corrected by the average efficiencies for the analysis cuts, and then by the average hardware and software thresholds.  Since the recoil-energy spectra and the former efficiencies are functions of $Q$-corrected recoil energy, before dividing out the latter efficiencies they are converted from $Y_{\mathrm{NR}}$-corrected to $Q$-corrected recoil energy by smearing with the ionization noise.

\subsection{\label{sec:4b}Backgrounds}

Although the recoil spectra resemble in shape the distributions expected for WIMP interactions, the events in the signal region are likely due to several types of unrelated background processes consisting of electron recoils, zero-ionization events, $^{14}$C contamination particular to Z6, and nuclear recoils from cosmogenic neutrons.  We will not subtract these events, but will accept them as candidates for the purpose of calculating upper limits on a WIMP signal; this is the most conservative treatment of these data.

\begin{figure}[t!]
\centering
\includegraphics[width=3.4in]{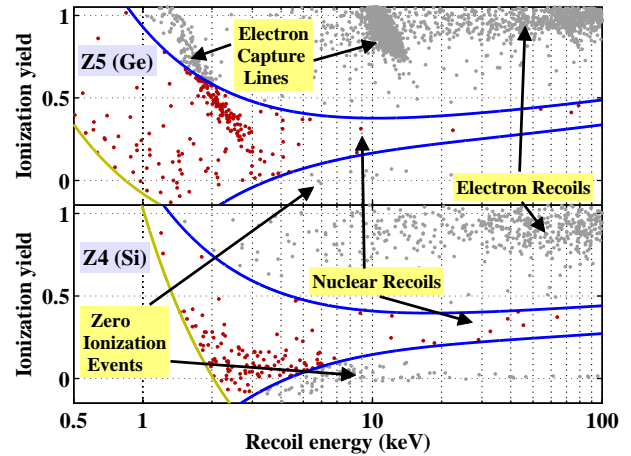}
\caption{(color online).  Ionization yield versus recoil energy for unvetoed single scatters passing the data-quality and fiducial-volume cuts (all dots) for representative Ge (top panel, Z5 6V) and Si (bottom panel, Z4 3V) WIMP searches.  Each detector's signal region is outlined in this plane by its nuclear-recoil band (blue/dark solid), phonon energy software threshold (yellow/light solid), and the extent of the horizontal axis.  Although these regions are partially cut off from above and below, all 202 Z5 6V and 130 Z4 3V WIMP candidate events (red/dark dots) can be seen.}
\label{fig:yplots}
\end{figure}

A few background populations are particularly evident in plots of ionization yield versus recoil energy.  The signal regions and candidate events for representative Ge and Si detectors are displayed in Fig.~\ref{fig:yplots}.  The most easily identified background is specific to the Ge detectors.  The distinct 1.3\,keV line between 1\,keV and 3\,keV in recoil energy accounts for a substantial number of the candidate events.  On average, the internal electron capture x-rays or Auger-electrons from the decays of $^{68}$Ge and $^{71}$Ge have unit ionization yield.  Due to the relatively low signal-to-noise ratio at these energies in both phonons and ionization, however, electronic noise induces a large tail of electron recoils to low ionization yield.  The feature is tilted with respect to the recoil-energy axis because of anticorrelation between the numerator and denominator of the ionization yield expression.  Leaked electron recoils from the 1.3\,keV line account for $\sim$20\% of the Ge candidates in the 3V WIMP search, and for approximately one-third (Z3) to one-half (Z2 and Z5) of the Ge candidates in the 6V WIMP search.  The fraction is greater for the 6V data because extensive \cf neutron calibrations were performed prior to this data period, enhancing the levels of $^{71}$Ge via thermal neutron capture.

Although more prominent for the Si detectors, both the Ge and Si detectors are afflicted by a distribution of events with nearly zero ionization yield.  Close examination reveals that these ``zero-charge'' events possess ionization signals indistinguishable from electronic noise.  They are otherwise normal events with recoil energies as large as 100 keV in some instances.  The distribution of zero-charge events for the Si detector in Fig.~\ref{fig:yplots} spans the full analysis energy range, and clearly crosses into the signal region for recoil energies $\lesssim$7\,keV, constituting the majority of this detector's candidate events. A possible explanation is that these events were the result of recoils that occurred near the side edges of the ZIP detectors, where electric field lines did not span the detectors' 1\,cm thicknesses.  Ionization occurring along field lines that terminate on a detector's edge rather than on an electrode is not properly drifted across the crystal, resulting in no signals in either the inner or outer ionization electrodes.  For recoil energies $\gtrsim$10\,keV, the zero-charge events' xy-positions can be reliably reconstructed, and tend to cluster near detector edges.  Past analyses of CDMS data have avoided zero-charge events by requiring WIMP candidates to have a minimum ionization energy, analogous to the phonon software thresholds described above.  Unfortunately, an ionization threshold would severely limit our WIMP detection efficiency for recoil energies $\lesssim$5\,keV.

The significantly higher event rate for Z6 is believed to be due to $^{14}$C surface contamination.  Prior to Run~21, during testing at one of the CDMS test facilities, Z6 was operated in close proximity to a detector that had been previously exposed to a $^{14}$C calibration source with faulty encapsulation, accidentally contaminating one of its surfaces with a low level of the isotope.  For this reason, Z6 was placed at the bottom of the detector tower with its contaminated surface facing away from the adjacent detector.  Beta decays of $^{14}$C produce electrons with an average energy of $\sim$50\,keV and a maximum energy of $\sim$156\,keV.  Beta radiation in this energy range will interact entirely within a ZIP detector's $\sim$10\,$\mu$m surface dead layer, where the charge collection efficiency is considerably reduced.  These events have reduced ionization yield, and populate the gap between the bands of electron and nuclear recoils when plotted in the fashion of Fig.~\ref{fig:yplots}, with a substantial number leaking into the signal region.  Although surface events can be rejected with high efficiency for recoil energies $\gtrsim$10\,keV through a combination of phonon and ionization pulse timing parameters, the near-threshold WIMP detection efficiency cannot be preserved.

Leakage of electron recoils into the nuclear-recoil band is a component of each detector's candidate events, although the source is usually Compton scatters of photons.  The discrimination based on ionization yield breaks down as the recoil energy decreases, until at the ``crossover energy'' the electron- and nuclear-recoil bands significantly overlap.  The crossover energy varies from $\sim$3\,keV (Z5 3V) to $\sim$7\,keV (Z6 6V).  More Neganov-Luke phonons are produced when the detectors are run at higher bias voltage, causing degradation in both the recoil-energy resolution and the yield-based discrimination.  Consequently, ZIP detectors perform better as low-threshold detectors with the lower 3V bias voltage.  Scaling calibration data from a \co source results in the estimate that only a few Compton electron-recoil events per detector leak into the nuclear-recoil band for recoil energies above the crossover energy.  We have not devised a reliable method of estimating the contribution of Compton electron-recoil leakage for recoil energies below the crossover energy.  We estimate that 10\% to 20\% of the WIMP candidates are actually electron recoils from Compton scatters.

The highest-energy signal events are largely due to the neutron background associated with the SUF's modest overburden.  Muons (and hadronic showers produced by them) occasionally broke apart nuclei in the rock surrounding the experiment, expelling high-energy neutrons with sufficient energy to punch through our shielding and create lower-energy neutron secondaries within the shielding materials capable of producing signals above threshold.  The expected rate of these unvetoed neutron interactions as a function of recoil energy was simulated for the Ge detectors (dashed curve in Fig.~2 in \cite{cdms2}).  With no efficiencies applied, the rate peaks at $\sim$0.1\,events\,kg$^{-1}$keV$^{-1}$d$^{-1}$ at the 0.5\,keV cutoff, and decays quasiexponentially to $\sim$0.003\,events\,kg$^{-1}$keV$^{-1}$d$^{-1}$ at 100\,keV.  Application of the Ge detectors' average detection efficiencies, followed by a scaling to the $\sim$71\,kg-days of Ge exposure, yields an expectation of $\sim$66 neutrons among the 1080 Ge candidate events.  

An accurate accounting of the contribution of each source of background to the total number of candidate events is difficult.  Most candidate events have low recoil energies for which our experimental variables are unable to differentiate among the various sources, particularly for the zero-charge and Compton electron-recoil leakage backgrounds.  An approximate tally of the percentage of candidate events due to each background source is listed in Table~\ref{tab:backgrounds} for the combined Ge and Si detector ensembles, where an ``other'' category is included to indicate the percentage of events that are not attributed to the background sources as described above.  The other events could be due to unidentified sources of background or WIMPs, but probably reflect systematic uncertainty in the estimates for the known background sources.

\begin{table}[t!]
\renewcommand{\arraystretch}{1.4}
\caption{Percentages of the 1080 Ge and 970 Si WIMP candidate events caused by each of the known background sources:  1)leaked electron recoils from the 1.3\,keV line; 2)zero-charge events occurring near detector edges; 3)leaked electron recoils due to beta decays of $^{14}$C embedded in the surface of the Si detector Z6; 4)leaked electron recoils due to Compton scatters of photons; and 5)nuclear recoils due to neutrons.  The final category labeled ``other'' indicates the percentage of events not attributed to the known background sources as estimated.} 
\label{tab:backgrounds}
	\begin{center}
		\begin{ruledtabular}
			\begin{tabular*}{3.4in}{@{\extracolsep{\fill}} l c c} 
				& \multicolumn{1}{c}{Ge (\%)} & \multicolumn{1}{c}{Si (\%)}\\ \hline 
				Electron Capture 1.3\,keV Line&32&0\\
				Zero-Charge Events&30{\textendash}40&30{\textendash}40\\
				$^{14}$C Contamination Betas&0&40\\
				Compton Photon Electron Recoils&10{\textendash}20&10{\textendash}20\\
				Cosmogenically-Induced Neutrons&6&2\\
				Other&2{\textendash}22&0{\textendash}18\\ 
			\end{tabular*}
		\end{ruledtabular}
	\end{center}
\end{table}

\subsection{\label{sec:4c}Exclusion limits}

The large uncertainties associated with our background sources preclude the subtraction of backgrounds.  We can claim no evidence of a WIMP signal, and proceed to calculate conservative limits under the assumption that the candidate events may constitute a WIMP signal.  We use the observed event rates to set upper limits on the WIMP-nucleon spin-independent elastic scattering cross section as a function of WIMP mass.  Two limits are calculated; the main result of this paper is a combined Ge and Si exclusion limit based upon all 2050 signal events, while a secondary result focuses on the Si data alone.

To calculate exclusion limits the observed event rates must be compared to a hypothetical WIMP model.  For convenient comparison with the results of other direct detection experiments, we work within the framework of the ``standard'' halo model described in \cite{halo1}, but normalized to a local WIMP density of 0.3\,GeV/cm$^{3}$.  We also conformed to the dark matter community's standard assumptions for the WIMP characteristic and mean Earth velocities of 220\,km/s and 232\,km/s, respectively~\cite{halo2}, while assuming the more recently estimated value of 544\,km/s for the galactic escape velocity~\cite{rave1}.  The effect of the galactic escape velocity on these results is discussed in more detail in Section~\ref{sec:4d}. 

Limits were calculated using a version of Yellin's ``optimum interval'' method~\cite{yellin1} that has been extended to accommodate high statistics~\cite{yellin2}.  For each choice of WIMP mass, the limit is essentially determined by a single energy interval for which the number of observed events is particularly low relative to the expected number of WIMP events.  An appropriate statistical penalty is applied for choosing the interval that sets the best limit, yielding a 90\% confidence level upper limit.  This method is especially effective at discriminating against backgrounds which are distributed differently from the expected signal.

To calculate a single exclusion limit, the data from the individual Ge and Si detectors have to be appropriately combined.  Traditionally, CDMS has combined the detector ensemble into a single averaged detector, where the individual-detector masses and efficiencies are averaged according to their exposures.  This ``averaged'' method for combining detectors makes use of the entire exposure, and comingles the candidate event energies for different detectors before forming the energy intervals required by the optimum interval method.  Figure~\ref{fig:nr_spectra} represents the averaged version of the candidate event data for this analysis.  The averaging method is appropriate when the detectors involved have approximately equal sensitivity to WIMP interactions, as was the case for previous CDMS WIMP-search results in which the analyses were either background-free or nearly so.

When the averaging technique is applied to detectors with variable event rates, the detectors with especially high event rates effectively pollute the lower-rate detectors by filling in the most sensitive intervals with a disproportionate number of events.  For this reason, we decided to adopt a novel ``serialization'' technique for combining the detector data.  Energy intervals are separately prepared for each detector in order to preserve the most sensitive intervals.  The intervals are then concatenated in an arbitrary order which, to avoid possible bias, was selected before the effect of the order was known.  We chose to place the 3V data before the 6V data, and then to order them according to their position within the detector tower (from top to bottom).  If the limit-setting intervals do not span multiple detectors, the order will not affect the result.  This technique allows the optimum interval method to calculate the limit from the best individual-detector energy intervals.  The resulting limit reflects only a fraction of the exposure, rather than the total exposure for the entire detector ensemble.  This is a trade-off we decided to accept before calculating the limits.  Each detector is clearly background-limited, particularly near threshold where our low-mass WIMP sensitivity resides.  Trading exposure for cleaner energy intervals should yield stronger limits for low masses.  In hindsight, this turned out to be true for WIMP masses less than 8\,\gev.  The serialization technique also allows for different detector types within the detector ensemble, providing a natural method for combining the Ge and Si data.

\begin{figure}[t!]
\centering
\includegraphics[width=3.4in]{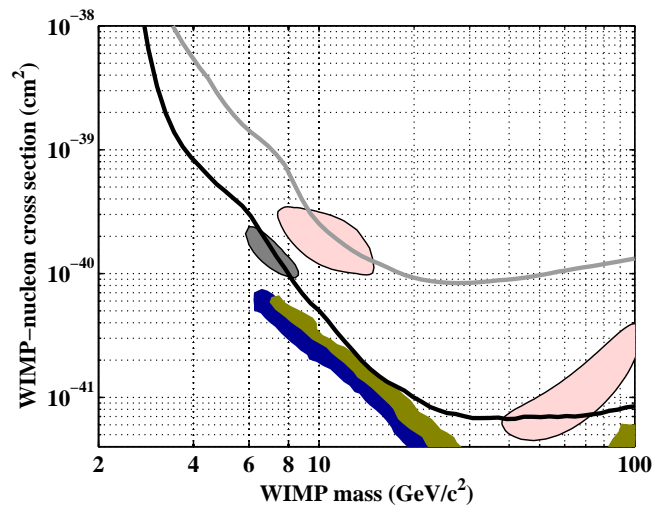}
\caption{(color online).  Comparison of 90\% confidence level upper limits from the combined Ge and Si (black/dark solid, our main result) and Si only (gray/light solid) data, with potential signal regions based on data from the DAMA/LIBRA~\cite{dama2,*dama3} and CoGeNT~\cite{cogent1} experiments.  The two (larger) oval-shaped filled regions (pink/light shaded) represent the DAMA/LIBRA annular modulation signal as interpreted by Savage \etal~\cite{savage1,*savage2,*savage3} (99.7\%\,C.L.), and include the effect of ion channeling as modeled by Bozorgnia \etal~\cite{channel1,*channel2}.  The (smaller) oval-shaped filled region is the 99\% (gray/medium shaded) confidence level signal region found by Hooper \etal's~\cite{hooper} simultaneous best fit to the DAMA/LIBRA and CoGeNT data.  The elongated filled regions are SUSY theory predictions by Bottino \etal~\cite{lsp3} for $\Omega_{\mathrm{WIMP}}<\Omega_{\mathrm{CDMmin}}$ (dark-yellow/medium shaded) and $\Omega_{\mathrm{WIMP}}\geq\Omega_{\mathrm{CDMmin}}$ (blue/dark shaded).  Our limits assume a galactic escape velocity of 544\,km/s~\cite{rave1}, while the potential signal regions are based on a value of 600\,km/s.}
\label{fig:LimitPlot1}
\end{figure}

To include the effect of non-zero energy resolution properly, expected WIMP rates were separately calculated for each detector and WIMP search (3V and 6V data) in a series of steps.  The limit was calculated for 75 WIMP masses between 1\,\gev and 100\,\gev.  At each mass, the halo model predicts the differential WIMP-nucleon scattering rate in terms of an ideal, perfect-resolution recoil energy (see Fig.~\ref{fig:WIMPrates} for example).  Each detector's ideal spectrum was then convolved with its $Y_{\mathrm{NR}}$-corrected recoil-energy resolution listed in Table~\ref{tab:resolution} (first two columns).  Recall that the hardware trigger and software phonon thresholds depend solely on the phonon signal.  The expected WIMP spectrum should therefore include noise from only the phonon channel when the hardware and software threshold efficiencies are applied.  After application of these phonon-only efficiencies, the spectrum was further smeared to include the electronic noise of the ionization channel via a second convolution with the quadrature difference between the $Q$-corrected (third and fourth columns of Table~\ref{tab:resolution}) and $Y_{\mathrm{NR}}$-corrected recoil-energy resolutions.  The threshold-reduced expected WIMP spectrum, in terms of $Q$-corrected recoil energy as measured by a ZIP detector, was then multiplied by the remaining analysis cut efficiencies, which either depend weakly on this energy estimator or are constant.  Finally each detector's doubly smeared and efficiency-reduced expected WIMP rate was scaled by the detector's mass and exposure (listed in Tables~\ref{tab:det_masses} and ~\ref{tab:candidates}).  The resulting distribution of expected events per $Q$-corrected keV for each detector was used to construct a cumulative probability that describes how likely it is that a WIMP interaction deposited an energy within each energy interval defined by that detector's candidate event energies.  The 10 (4) individual probability distributions were serialized as described above, and a combined Ge and Si (Si only) optimum interval upper limit was calculated at each WIMP mass.  Note that although particular care was taken to include the effect of non-zero energy resolution for this result, it would have made only a trivial contribution to the sensitivities of previous CDMS results due to their higher thresholds.

\begin{figure}[t!]
\centering
\includegraphics[width=3.4in]{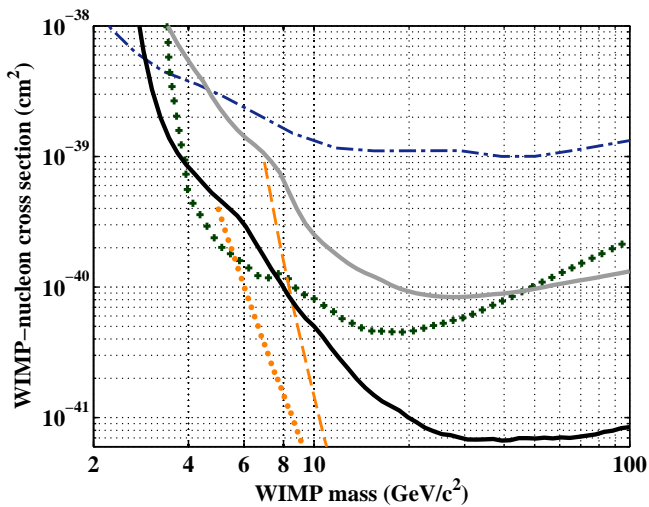}
\caption{(color online).  Comparison of 90\% confidence level upper limits from the combined Ge and Si (black/dark solid, our main result) and Si only (gray/light solid) data, with those from XENON100 with constant (orange/light dotted) or decreasing (orange/light dashed) scintillation efficiency extrapolations at low energy~\cite{xenon2}, CoGeNT~\cite{cogent1} ($+$), and CRESST~\cite{cresst1} (blue/dark dash-dotted).  Our limits (and XENON100's) assume a galactic escape velocity of 544\,km/s~\cite{rave1}, while the CRESST limit uses 650\,km/s and the CoGeNT limit uses 600\,km/s.  See also Fig.~\ref{fig:LimitPlot2} in which limits for other escape velocities are compared.}
\label{fig:LimitPlot3}
\end{figure}

The resulting exclusion limits are shown in Fig.~\ref{fig:LimitPlot1}, and partially exclude parameter space associated with potential signals from the DAMA/LIBRA~\cite{savage1,*savage2,*savage3,dama2,*dama3} and CoGeNT~\cite{cogent1} experiments.  The former includes the (small) effect of ion channeling as modeled by Bozorgnia \etal~\cite{channel1,*channel2}.  The combined Ge and Si limit cuts through the middle of the 99\% confidence level signal region associated with Hooper \etal's~\cite{hooper} simultaneous fit to the DAMA/LIBRA and CoGeNT data, and excludes new parameter space for WIMP masses between 3\,\gev and 4\,\gev.  Our limits are compared to those from other experiments with strong low-mass WIMP sensitivity in Fig.~\ref{fig:LimitPlot3}.  The additional information required to reproduce our results can be found in \cite{[{See \href{http://link.aps.org/supplemental/10.1103/PhysRevD.82.122004}{EPAPS supplementary material} for the WIMP candidate event energies and additional efficiencies required to reproduce our upper limits.}]epaps}, and includes the efficiencies and candidate event energies for each detector and voltage bias.

A limit based on the Ge data alone was also calculated, and found to be almost equal to the combined Ge and Si limit (the difference is imperceptible in Fig.s~\ref{fig:LimitPlot1} and ~\ref{fig:LimitPlot2}).  The serialization technique sets the limit according to the most sensitive detectors.   The Ge detectors were more sensitive across the range of WIMP masses considered, yielding a combined Ge and Si limit that is almost the same as the Ge only limit.  We decided beforehand that the combined detector limit would be our main result, but that we would publish the Ge only and Si only limits as well.  The limit for low masses ($\lesssim$6\,\gev) was determined by Z2 3V and 6V intervals, while a combination of the Z5 3V, Z5 6V, and Z2 3V data set the limit for WIMP masses $\gtrsim$10\,\gev.  The sensitivity in the intervening mass range was dominated by the Z3 3V data.  Each Si detector contributes to the Si only sensitivity, with the Z6 data providing the best limit for masses $\lesssim$4\,\gev, and the Z4 data for larger WIMP masses.  None of the intervals chosen by this serialized implementation of the optimum interval method spanned more than one detector, confirming that the order of the serialization is unimportant.

\subsection{\label{sec:4d}Systematic studies}

\begin{figure}[t!]
\centering
\includegraphics[width=3.4in]{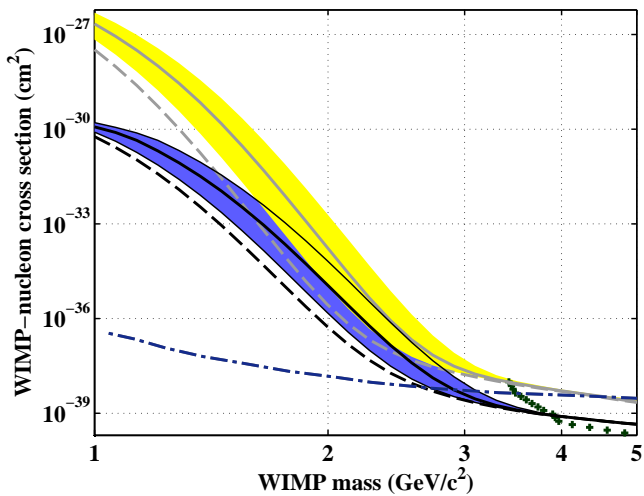}
\caption{(color online).  The 90\%\,C.L. upper limits from the combined Ge and Si and Si only data for WIMP masses down to 1\,\gev and an assumed galactic escape velocity of 544\,km/s~\cite{rave1} (black/dark solid and gray/light solid, respectively), compared to limits from the same data for an escape velocity of 650\,km/s~\cite{halo2} (black/dark dashed and gray/light dashed, respectively).  The ranges of combined Ge and Si (blue/dark shaded) and Si only (yellow/light shaded) limits corresponding to the 90\%\,C.L. range of escape velocities found by Smith \etal~\cite{rave1}, and the limits from the CRESST~\cite{cresst1} (blue/dark dash-dotted, 650\,km/s) and CoGeNT~\cite{cogent1} ($+$, 600\,km/s) experiments are shown as well.}
\label{fig:LimitPlot2}
\end{figure}

A recent study by Smith \etal~\cite{rave1} analyzed the highest velocity halo stars in an early release of data from the RAVE survey~\cite{rave2}, and derived a loose but convincing constraint on the galactic escape velocity.  At the 90\% confidence level, their preferred range of escape velocities is 498\,km/s to 608\,km/s, with a median value of 544\,km/s.  Though based on an otherwise identically normalized standard halo model, the results for the other experiments shown in Figs.~\ref{fig:LimitPlot1} and~\ref{fig:LimitPlot3} assume a variety of galactic escape velocities which are generally larger.  The CRESST~\cite{cresst1} limit is based on a much larger 650\,km/s value, while the DAMA/LIBRA and CoGeNT results (including the Hooper \etal region) use 600\,km/s.  Although we have chosen to adopt the more recent measurement of 544 km/s for this paper, additional limits were calculated to see the effect of the larger, 650\,km/s escape velocity.  There is very little visible difference in the main result for WIMP masses larger than 4\,\gev, while the Si only limit assuming the lower escape velocity is as much as 20\% weaker for WIMP masses between 6\,\gev and 10\,\gev.  The effect of changing the escape velocity becomes important for masses less than 4\,\gev.  Limits based on the 650\,km/s and 544\,km/s galactic escape velocities for WIMP masses down to 1\,\gev are compared in Figure~\ref{fig:LimitPlot2}.  Regions corresponding to Smith \etal's 90\% confidence interval are plotted as well.  The effect is nearly maximal for a 2\,\gev WIMP mass, where the combined Ge and Si limits span over an order of magnitude, and the Si only limits span over two orders of magnitude.

The detection efficiencies associated with each detector's hardware and software phonon thresholds, and analysis cuts, were estimated directly from data.  In each case, the estimates are based on a finite number of events, resulting in statistical uncertainties.  These uncertainties have a small (but noticeable) effect on our experimental sensitivity for WIMP masses less than 6\,\gev.  The combined Ge and Si exclusion limit for a 5\,\gev WIMP, for example, is $\sim$20\% weaker when the 90\%\,C.L. lower limit efficiencies (1.28$\sigma$ less than the mean) are substituted for the best-fit mean efficiencies.  The effect is maximal for a 1\,\gev WIMP, where the same WIMP exclusion limit is a factor of 5 weaker.  Again, we decided to adopt a conservative approach and employ the 90\%\,C.L. (statistical) lower limit efficiencies in all our WIMP exclusion limit calculations.  The best-fit mean efficiencies are compared to these reduced efficiencies in Fig.~\ref{fig:reduced_eff} for representative Ge and Si detectors.  All figures and tables in this paper preceding Figs.~\ref{fig:LimitPlot1}\textendash\ref{fig:LimitPlot2} reflect the best-fit mean efficiencies where applicable, while Figs.~\ref{fig:LimitPlot1}\textendash\ref{fig:LimitPlot2} are based on reduced detection efficiencies such as those shown in Fig.~\ref{fig:reduced_eff}.

The combined Ge and Si exclusion limit was also tested for sensitivity to uncertainties in the Ge detectors' electron-recoil energy scales.  Each Ge ZIP's $Y_{\mathrm{ER}}$-corrected and $Q$-corrected recoil energy scales for electron recoils were checked for the entire 3V and 6V WIMP searches using the lines described previously.  The three peaks' mean energies were estimated with Gaussian fits relative to energy scales originally derived from short \cs calibrations.  Deviations between the observed and expected peak energies were generally within statistical fluctuations.  A few detectors, however, exhibited sizable phonon signal overestimates for the 1.3\,keV line.  The worst case was for the Z2 6V data, where the $Y_{\mathrm{ER}}$-corrected recoil energy was over measured by $\sim$6\%, and the $Q$-corrected recoil energy was over measured by $\sim$16\%.  The latter is worse because it includes an ionization signal under-measurement in conjunction with a phonon signal over-measurement.  An \textit{ad hoc} recalibration of the Ge recoil-energy estimators was performed and the WIMP exclusion limits were recalculated.  Because the energy scale discrepancies mostly involve recoil-energy overestimates, the combined Ge and Si limit based on the recalibrated energy scales is slightly stronger.  The limit is mostly insensitive to the recalibration, however, with the difference in limits only becoming visible by eye for the lowest WIMP masses.  The effect is maximal for a WIMP mass of 1\,\gev, where the recalibrated limit is a factor of $\sim$2 stronger.  The final limits shown in Figs.~\ref{fig:LimitPlot1}\textendash\ref{fig:LimitPlot2} are based on the original Ge detectors' energy scales, a slightly conservative choice that avoids the systematic uncertainty associated with the \textit{ad hoc} recalibration.

It is easy to imagine that our low-mass WIMP sensitivity might be heavily reliant on the lowest-energy trigger efficiency, and on our inclusion of the effect of non-zero energy resolution near threshold.  If true, then relatively modest unknown systematic uncertainties in the hardware trigger efficiencies could lead to large uncertainties in the WIMP exclusion limits.  The combined Ge and Si limit was tested for sensitivity to potential trigger efficiency systematic uncertainty in two ways.  First, each detector's hardware trigger efficiency was set equal to zero wherever it drops below 10\%, eliminating the lowest-energy detection efficiency.  The limit based on the reduced trigger efficiencies is reassuringly not much weaker than the limit using the full efficiencies.  The difference is maximal at 1\,\gev, where the reduced-efficiency limit is $\sim$3\% weaker.  Second, the sensitivity to possible systematic uncertainty associated with the inclusion of near-threshold energy resolution was tested by smearing the hardware trigger efficiencies by 50\% of the $Y_{\mathrm{NR}}$-corrected recoil-energy resolutions ($\sim$50\,eV for Ge, and $\sim$100\,eV for Si) before applying them in the limit calculation described above.  This addresses a concern that there might be an unknown extra resolution factor included in the hardware trigger efficiency estimates due to the difference between the phonon signal that triggers the experiment and the offline reconstructed phonon energy.  It would actually be more appropriate to deconvolve the trigger efficiencies, but the reverse operation is easier to implement, and although it will give a stronger rather than weaker limit, the size of the effect should be similar.  The resulting limit is indeed stronger, but only slightly so.  The size of the effect reaches a maximum for a 1\,\gev WIMP mass, where the difference is $\sim$5\%.  Our low-mass WIMP sensitivity appears to be relatively insensitive to systematic uncertainties in the hardware trigger efficiencies.

\section{\label{sec:5}Conclusions}

Within the dark matter community, possible signals from the DAMA/LIBRA and CoGeNT experiments have aroused renewed interest in relatively low-mass WIMPs.  Despite substantial backgrounds, CDMS ZIPs are excellent low-threshold detectors.  The analysis presented here is background-limited and conservative, yet sets competitive limits for low WIMP masses on the spin-independent WIMP-nucleon elastic scattering cross section.  Under standard halo assumptions, we partially exclude the parameter space favored by interpretations of the DAMA/LIBRA annual modulation and CoGeNT excess as WIMP signals.  Ignoring the effect due to uncertainty related to the galactic escape velocity, the limits are robust for WIMP masses greater than $\sim$2\,\gev.  

The methods developed for this low-threshold analysis are not limited to the Run~21 shallow-site data.  An upcoming analysis based on CDMS II data taken at our deep site in the Soudan Mine employs a similar but lower-background analysis technique, \ and achieves similar recoil-energy thresholds for a significantly greater exposure.  We have also experimented with operating CDMS detectors at very high bias voltages ($\sim$100\,V), sacrificing our two-channel discrimination in favor of ultra-low energy thresholds ($\sim$50\,eV) via Neganov-Luke phonon amplification~\cite{cdmslite}.  The CDMS II data are a rich source of information for probing low-mass WIMP parameter space.

\begin{acknowledgments}

The CDMS collaboration gratefully acknowledges the contributions of numerous engineers and technicians; we would like to especially thank Judith Alvaro-Dean, Jim Beaty, Sam Burke, Daniel Callahan, Pat Castle, John Emes, Merle Haldeman, David Hale, Michael Hennessey, Wayne Johnson, Jim Perales, Garth Smith and Astrid Tomada.  Additionally, we would like to thank former CDMS collaborators for their contributions to the successful completion of this work, including Long Duong, Jochen Hellmig, Al Lu, John Martinis, Thushara Perera, Maria Perillo Isaac, Ron Ross, Tony Spadafora, and John-Paul Thompson.  This work is supported in part by the National Science Foundation (Grant Nos. AST-9978911, \ PHY-0542066, \ PHY-0503729, PHY-0503629, \ PHY-0503641, PHY-0504224, \ PHY-0705052, \ PHY-0801708, PHY-0801712, PHY-0802575, PHY-0855525, and PHY-9722414), by the Department of Energy (Contracts DE-AC03-76SF00098, DE-FG02-91ER40688, DE-FG02-92ER40701, DE-FG03-90ER40569, and DE-FG03-91ER40618), by the Swiss National Foundation (SNF Grant No. 20-118119), and by NSERC Canada (Grant SAPIN 341314-07).

\end{acknowledgments}

\bibliography{lowTr21}

\bibliographystyle{apsrev4-1}

\end{document}